\documentclass[a4paper,twoside]{article}
\usepackage{subfigure}
\usepackage{calc}
\usepackage{amssymb}
\usepackage{amstext}
\usepackage{amsmath}
\usepackage{amsthm}
\usepackage{multicol}
\usepackage{pslatex}
\usepackage{apalike}
\usepackage{isabelle}
\usepackage{isabellesym}
\usepackage{pifont} 
\usepackage{stmaryrd} 
\usepackage{url}
\usepackage{html}
\usepackage{mdwlist}
\usepackage{xspace}
\usepackage{focus}
\usepackage[usenames,dvipsnames]{xcolor} 

\usepackage{SCITEPRESS}      

\begin{document}  

\newcommand{\isah}{Isabelle/HOL\xspace}
\newcommand{\tlangle}{\text{\textlbrackdbl}}
\newcommand{\trangle}{\text{\textrbrackdbl}}
\newcommand{\ntlangle}{{\normalfont\text{\textlbrackdbl}}}
\newcommand{\ntrangle}{{\normalfont\text{\textrbrackdbl}}}

\newcommand{\secret}{\emph{Secret}\xspace}
\newcommand{\keys}{\emph{Keys}\xspace}
\newcommand{\keysecret}{\emph{KeysSecrets}\xspace}
\newcommand{\channels}{\emph{Channels}\xspace}
\newcommand{\vars}{\emph{Var}\xspace}
\newcommand{\enc}{\emph{Enc}\xspace}
\newcommand{\Rel}{\mathsf{relevant}}
\newcommand{\Enc}{\mathsf{enc}}
\newcommand{\Sig}{\mathsf{sig}}

\renewcommand{\fti}[2]{\ensuremath{#1}^{\ensuremath{#2}}_{\textsf{ft}}}
\renewcommand{\tempty}[1]{\ti{#1}{t} = \nempty}
\renewcommand{\tnempty}[1]{\ti{#1}{t} \neq \nempty}
\renewcommand{\titrue}[1]{\ti{#1}{t} = \angles{\ntrue}}
\renewcommand{\tifalse}[1]{\ti{#1}{t} = \angles{\nfalse}}

\newcommand{\ist}[1]{{\isastyle #1}}
\newcommand{\tbf}[1]{\textbf{#1}}
\newcommand{\tit}[1]{\textit{#1}}
\newcommand{\msfn}[1]{\textsf{\small#1}}
\newcommand{\msf}[1]{\textsf{\footnotesize#1}}
\newcommand{\mbsf}[1]{\textsf{\textbf{\footnotesize#1}}}
\newcommand{\mblue}[1]{\textcolor{blue}{#1}}
\renewcommand{\nint}[2]{\ensuremath{(#1\,{\rhd}\,#2)}}
\renewcommand{\angles}[1]{\ensuremath{\langle #1\rangle}}
\renewcommand{\nempty}{\angles{}}
\renewcommand{\ti}[2]{\ensuremath{#1}^{\ensuremath{#2}}}
\renewcommand{\Nti}[1]{\textsf{Nti}(\ensuremath{#1})}
\renewcommand{\msg}[2]{\textsf{msg}_{#1}(#2)} 
\renewcommand{\ndisjoint}[1]{\textsf{disjoint}(#1)} 
\renewcommand{\semantics}[1]{\textlbrackdbl\xspace#1\xspace\textrbrackdbl\xspace}
\renewcommand{\semanticsm}[1]{\ensuremath{[\![\ #1\ ]\!]}} 

\newcommand{\snd}[1]{\textsf{snd}.{#1}}
\newcommand{\trd}[1]{\textsf{trd}.{#1}}
\newcommand{\eout}[2]{{#1}^\textsf{eout}(#2)}
\newcommand{\eouts}[3]{{#1}^\textsf{eout}_{#3}(#2)}
\newcommand{\ine}[2]{{#1}^\textsf{ine}(#2)}
\newcommand{\ines}[3]{{#1}^\textsf{ine}_{#3}(#2)}
\newcommand{\lsecret}[2]{{#1}^\textsf{leak}(#2)}
\newcommand{\lsecretc}[3]{{#1}^\textsf{leak}_{#3}(#2)}
\newcommand{\secrecy}[2]{{#1}^\textsf{secr}(#2)}
\newcommand{\secrecyc}[3]{{#1}^\textsf{secr}_{#3}(#2)}
\newcommand{\knows}[2]{\textsf{knows}^{#1}(#2)}
\newcommand{\know}[2]{\textsf{know}^{#1}(#2)}
\newcommand{\knowst}[2]{\textsf{knows}^{#1}_{#2}}
\newcommand{\tkn}[2]{t_\textsf{know}^{#1}(#2)}
\newcommand{\tkns}[2]{t_\textsf{knows}^{#1}(#2)}
\newcommand{\knowss}[3]{\textsf{knows}^{#1}_{#3}(#2)}
\newcommand{\tknc}[3]{t_\textsf{knowS}^{#1}_{#3}(#2)}
\newcommand{\tkncs}[3]{t_\textsf{knowsS}^{#1}_{#3}(#2)}
\newcommand{\ntypeKS}[1]{\textsf{typeKS}(#1)}
\newcommand{\ntypeCExp}[1]{\textsf{typeCExp}(#1)}
\newcommand{\ntypeEnc}[1]{\textsf{typeEnc}(#1)}
\newcommand{\subcomp}[1]{\textsf{subcomp}(#1)}
\newcommand{\sti}[2]{\ensuremath{#1}^{\ensuremath{#2}}_{\textsf{snd}}}
\newcommand{\tti}[2]{\ensuremath{#1}^{\ensuremath{#2}}_{\textsf{trd}}}

\newtheorem{mytheorem}{Theorem}
\newtheorem{mydef}{Definition}
\newtheorem{myprop}{Proposition}
\newtheorem{ax}{Axiom}
\newcommand{\focust}{\textsc{Focus}$^{ST}$}

\pagestyle{myheadings}
\markboth{PREPRINT}{PREPRINT}

\title{Focus$^{ST}$ Solution for Analysis of Cryptographic Properties} 


 \author{\authorname{Maria Spichkova, Radhika Bhat}
 \affiliation{School of Science, RMIT University, Melbourne, Australia} 
 \email{\{maria.spichkova, s3703794@student.rmit.edu.au\}@rmit.edu.au
 }}

\abstract{
To analyse cryptographic properties of distributed systems in a systematic way, 
a formal theory is required. 
In this paper, we present a theory that allows (1) to specify distributed
systems formally, (2) to verify their cryptographic wrt. composition properties, and
(3)  to demonstrate the correctness of
syntactic interfaces for specified system components automatically.  
To demonstrate the feasibility of the approach we use a typical example from the domain of crypto-based systems: a variant of the Internet security protocol TLS. A security flaw in the initial version of TLS specification was revealed using a semi-automatic theorem prover, Isabelle/HOL.     
}

\onecolumn 
\maketitle 
\normalsize \vfill

\section{\uppercase{Introduction}}

\noindent 
Systems are often specified and implemented following the modularity principle: 
a number of separate components are combined together to build the desired system. 
This usually leads to the question on how to derive the system properties 
from the properties of its components. 
In the case of crypto-based systems, the most important and the most difficult question is to derive which of the cryptographic (security/secrecy) properties the composed system will have. 
Thus, a formal theory is required to not only specify 
systems and their cryptographic formally, but also to analyse them. 
As the paper-and-pencil proofs are not enough for this case, applying 
theorem provers or model checkers is necessary to have semi-automated solutions.  
 
In this paper, we discuss a formal theory for specification and verification of security-critical systems and their cryptographic properties.
The focal point of this approach is readability of formal specifications as well as the composition of components and their properties. 
The modelling language we use in our approach, \focust,  allows us to create concise but easily understandable specifications
and is appropriate for application of the specification and proof methodology presented in our previous works.

\focust\cite{spichkova2014modeling,spichkova2016spatio} is based on human factor analysis within formal methods to offer  more readable specifications. 
 The \focust\ language was inspired by \Focus~\cite{focus},
a framework for formal specification and development of interactive systems. 
In both languages, specifications are based on the notion of \emph{streams} that represent a 
communication history of  a \emph{directed channel} between components.  
However, in the original \Focus\  input and output streams of a component are mappings 
 of natural numbers $\Nat$ to the single messages,  
where a \focust\  stream %
 is a mapping from $\Nat$ to lists of messages  within the corresponding time intervals.  
The \focust\ specification layout also differs from the original one: it 
is based on human factor analysis within formal methods~\cite{hffm_spichkova,spichkova2013design}.  

 This theory is a result of optimization and extension (on the verification as well as on the specification level) of the draft ideas presented in technical reports~\cite{sj_TB08,spec_verif_crypto}. 
Similar to our previous work on formal specification of security-critical systems, 
we apply the proposed theory on a typical example from the domain of crypto-based systems: a variant of the Internet security protocol TLS \cite{tls}. We also discuss the differences to the corresponding \Focus specifications, especially focusing on the readability aspects. 
Using the extended approach with \focust, we can demonstrate a security flaw in the protocol
and show how to prove security properties of a corrected version.  
We also can apply to the \focust\ solution the verification methodology \emph{Focus on Isabelle} \cite{spichkova}, which allows verification using an interactive semi-automatic Higher-Order Logic theorem prover \isah. 
The corresponding proofs are presented in the Archive of Formal Proofs~\cite{spichkova2014compositional}.

\section{\uppercase{Related Work}}
\label{sec:related}

\noindent 
Security is critical to the development of software systems in many application areas. Thus,  there also are many approaches on developing of such systems. 
A brief survey of software engineering techniques for computer security can be found in~\cite{security_roadmap}. 
The closest  to our work in this field is the approach for secure software engineering using the CASE tool AutoFocus presented in~\cite{wimmel}: it uses a modelling tool based on the restricted part of the \Focus specification language, but do not cover the aspects of verification and properties composition, which we concentrate on. 

There are also many papers on verifying cryptographic protocols~\cite{Paulson:1998:IAV:353677.353681,Meadows:2001,Ryan:2000:MAS:1407727}. 
In comparison to them, we do not focus in our work on the protocol verification itself, but use the TLS protocol as a case study to show the advantages of our theory.

A large number of the  approaches focus on model-based development of security-critical
systems, cf. e.g.,  \cite{AlaHafBre07,DBLP:conf/icse/WhittleWH08}, however,
a correct composition of system specifications or system models and, in particular, deriving the properties of a composed system is treated as one of the most difficult objective~\cite{broy_refinements,DBLP:conf/ecmdafa/BezivinBFGJKKP06,DBLP:conf/fm/BrunetCU06} independently which kinds of system properties are discussed.  
Moreover, dealing with the composition of security-critical components and their security properties we get even more complex and costly task, to solve it we need to develop an appropriate theory of composition which allows a formal verification/derivation of  system properties in addition to a readable specification of system components. 

An approach on the verification of equivalence properties was introduced in \cite{chadha2012automated}. For security analysis of padding-based encryption schemes 
was presented in \cite{barthe2013fully}. 
There were also a number of approaches applied symbolic analysis of security protocols. 
For example, approach based on multi-set rewriting systems and first-order logic was presented in \cite{schmidt2012automated}. An approach presented in \cite{meier2013tamarin}, 
focuses on efficient deduction and equational reasoning, and introduces the corresponding 
 TAMARIN prover. 
 Model Checking solutions are also very popular, cf. e.g., 
\cite{permpoontanalarp2010fly}.
 Comparative Analysis of 15 Model Checking tools for security protocol verification was presented in \cite{patel2010comparative}, proposing the Scyther and AVISPA tools as mostly suitable for the purpose.  
 In our case, the approach is supported by Isabelle/Isar theorem prover for higher-order logic.


\section{\uppercase{Background: Focus$^{ST}$}}
\label{sec:focust}

A system in \Focus\ and \focust\  is represented by its components that are 
connected by communication lines called \emph{channels}, 
and are described in terms of its input/output
behaviour. The components can interact and also work independently of each other.
A specification can be elementary or composite, where composite specifications are
built hierarchically from the elementary ones.  
In both languages, any specification characterizes the relation between the
\emph{communication histories} for the external \emph{input} and \emph{output channels}, and the formal meaning of a specification is exactly this external \emph{input/output relation}. 
 
For any set of messages $M$, $M^\omega$ denotes the set of all streams,
$M^\infty$ and $M^*$ denote  the sets of all infinite and all finite
streams respectively, $M^{\underline{\omega}}$ denotes the set of all 
timed streams, $M^{\underline{\infty}}$ and $M^{\underline{*}}$ denote the sets of 
all infinite and all finite timed streams respectively.

The \Focus\ and \focust\ specifications can be structured into a number of
formulas each characterizing a different kind of properties. 
These languages support a variety of \emph{specification styles} which describe system
components by logical formulas or by diagrams and tables representing logical formulas. 
The most general style in \Focus\ and \focust\ is an Assumption/Guarantee style, where a component is specified in terms of an assumption and a guarantee: whenever input from the environment behaves in accordance
with the assumption $\msfn{asm}$, the specified component is required to fulfill the
guarantee $\msfn{gar}$.

We specify the semantics of a \emph{composite} component $S = S_1 \otimes \dots \otimes S_n$ 
as defined in~\cite{focus}:

\begin{equation}
		\semanticsm{S} \ndef 
		\exists l_S \in L_S:~ \bigwedge^n_{j=1} \semanticsm{S_j} 
\end{equation}
where $l_S$ denotes a set of \emph{local streams} and 
$L_S$ denotes their corresponding types, $\semanticsm{S_j}$  denotes semantics of the specification $S_j$, $1 \le j \le n$, which is a specification of subcomponent of $S$.

The collection of \focust operators over timing aspects and their properties specified and verified using the theorem prover Isabelle is presented in the Archive of Formal Proofs \cite{spichkova2013stream}.
In this work we focus on modelling of security aspects and the corresponding properties of composition. 
Before introducing the new concepts, we would like to mention very shortly a small number of operators we used in the paper:\\
An empty stream is represented by $\nempty$.\\ 
$\angles{x}$ denotes the one element stream consisting of the element $x$.\\
$\#s$ denotes the length of the stream $s$.\\
$i$th time interval of the stream $s$ is represented by $\ti{s}{i}$.\\
$\msg{n}{s}$ denotes a stream $s$ that can have at most $n$ messages at each time interval. 


\section{\uppercase{Secrecy}}
\label{sec:secrecy}

\noindent 
In this section we introduce a \focust\ formalization of security properties of data secrecy, 
corresponding definitions, and a number of abstract data types used in this formalization. 
This formalization yields a basis for verification in the theorem prover Isabelle/HOL, technical details of verification and the corresponding proofs are presented in~\cite{spec_verif_crypto}).

We assume here disjoint sets 
$Data$ of data values, \secret~of unguessable values, and \keys~of cryptographic  keys. 
Based on these sets, we specify the sets
\emph{EncType} of \emph{encryptors} that may be used for encryption or decryption, $CExp$ of closed expressions, and \emph{Expression} 
of expression items:
\[
	\begin{array}{lcl}
	KS & \ndef & \keys \cup \secret 
	\\
	EncType & \ndef & \keys \cup \vars
	\\
	CExp & \ndef & Data \cup \keys \cup \secret 
	\\ 	
	Expression & \ndef & Data \cup \keys \cup \secret \cup \vars 
	\end{array}
\]

\noindent%
Below, we will treat an \emph{expression} (that can for example be sent as
an argument of a message within the distributed system)
as a finite sequence of {expression} items.  
$\nempty$ then denotes  an empty expression.

The decryption key corresponding to an encryption key $K$ is written as $K^{-1}$. 
In the case of asymmetric encryption, the encryption key $K$ is public, and the decryption key $K^{-1}$ secret.
For symmetric encryption, $K$ and $K^{-1}$ coincide. 
For the encryption, decryption, signature creation and signature verification functions 
we define only their signatures and general axioms, because in order to
reason effectively, we view them as abstract functions and abstract from
their bit-level implementation details, following the usual Dolev-Yao approach
to crypto-protocol verification \cite{DY83}:
\[\begin{array}{l}
Enc,~Decr,~Sign,~Ext ::\\ EncType \times \nfst{Expression} \to  \nfst{Expression}
\\
\forall e \in Expression:
	 Ext(K, Sign(K^{-1}, e)) = e
	\\
	Decr(CKey^{-1}, Enc(CKey, e)) = e
	\end{array}
\]

\noindent
We denote by $\sse{K_P}{\keys}$ and $\sse{S_P}{\secret}$ the set of private keys  of a component $P$ and 
the set of unguessable values used by a component $P$, respectively.

We assume in our specification that the composition of components has a number of general properties which 
sometimes seem to be obvious, but for a formal representation is essential to mention these properties explicitly either we can't (edit:cannot) make the proofs in a correct way.

The sets of private keys and unguessable values used by a composed component 
$C = C_1 \otimes \dots \otimes C_n $ must be defined by union of corresponding sets.

(1) If $xb$ is a private key of the composed component $C$, then this key must belong to the set of private keys of one subcomponents of $C$:\\
$ 
C = C_1 \otimes \dots \otimes C_n \wedge xb \in K_C  \to \exists i \in [1..n].~ xb \in K_{C_i}
$
  
(2) 
If $xb$ is an unguessable value used by the composed component $C$, then this value must belong to the set of unguessable values used by one subcomponents of $C$:\\
$
C = C_1 \otimes \dots \otimes C_n \wedge xb \in S_C  \to \exists i \in [1..n].~ xb \in S_{C_i}
$
 
(3)
If $xb$ is a private key of one subcomponents of the composed component $C$, then this key must belong to the set of private keys of $C$:\\
$
C = C_1 \otimes \dots \otimes C_n \wedge 1 \le i \le n \wedge xb \in K_{C_i}  \to  xb \in K_C  
$

(4)
If $xb$ is an unguessable value used by one subcomponents of the composed component $C$, then this value must belong to the set of unguessable values used by  $C$:\\
$
C = C_1 \otimes \dots \otimes C_n \wedge 1 \le i \le n \wedge xb \in S_{C_i}  \to  xb \in S_C  
$
 
(5)
If $xb$ does not belong to the set of private keys and unguessable values of any subcomponent of $PQ = P \otimes Q$, then $xp$ does not belong to  the set of private keys and unguessable values  of $PQ$:\\
$
PQ = P  \otimes Q \wedge  xb \notin KS_{P}  \wedge  xb \notin KS_{Q}  \to  xb \notin KS_{PQ}
$
 
(6)
If a channel $x$ belongs to the set of input (output) channels of the composition   $PQ = P  \otimes Q$ for any two components $P$ and $Q$,
then this channel must belong to the set of input (output) channels of $P$ or $Q$:
\[ \begin{array}{l}
x \in i_{P  \otimes Q}  \to  x \in i_P \vee x \in i_Q
\\
x \in o_{P  \otimes Q}  \to  x \in o_P \vee x \in o_Q
\end{array}
\]

For the collection of  the theorems and prepositions on the input/output properties proven in \isah (more than 50 \isah lemmas) we would like to refer to~\cite{spichkova2014compositional}.

\subsection{Knowledges of an Adversary}
\label{sec:knows}

An (\emph{adversary}) component $A$ knows a secret $m \in KS$, $m \notin KS_A$ (or some secret expression $m$, $m \in \nfst{(Expression\setminus{KS_A})}$), if
\begin{itemize*}
	\item %
	$A$  may eventually get the secret $m$,
	\item 
	$m$ belongs to the set $LS_A$ of its local secrets, 
	\item %
	$A$ knows a one secret $\angles{m}$,
	\item %
	$A$ knows some list of expressions $m_2$ which is an concatenations of $m$ and some list of expressions $m_1$,
	\item %
	$m$ is a concatenation of some secrets $m_1$ and $m_2$ ($m = m_1 \nconc m_2$), and $A$ knows both these secrets,
	\item %
	$A$ knows some secret key $k^{-1}$ and the result of the encryption of the $m$ with the corresponding public key,
	\item %
	$A$ knows some public key $k$ and the result of the signature creation of the $m$ with the corresponding private key,
	\item %
	$m$ is an encryption of some secret $m_1$ with a public key $k$, and $A$ knows both $m_1$ and $k$,
	\item %
	$m$ is the result of the signature creation of the $m_1$ with the key $k$, and $A$ knows both $m_1$ and $k$.
\end{itemize*}
Formally, we define this term  by mutually recursive predicates $\know{A}{k}$ (for the case of a single secret $m$)   
and $\knows{A}{k}$ (for the case when expression (or list) $k$, containing a secret) respectively.
\[
\begin{array}{l}
\textsf{know}^{A} \in KS\setminus{KS_A} \to \Bool\\
\know{A}{m}  \ndef \ine{A}{m} ~\vee ~ m \in LS_A
\end{array}
\]
\[
\begin{array}{l}
\textsf{knows}^{A} \in \nfst{(Expression\setminus{KS_A})} \to \Bool
\\
\knows{A}{m}  \ndef\\
(\exists m_1: m = \angles{m_1} ~\wedge~ \know{A}{m_1})\\
~~\vee\\
(\exists m_1, m_2: (m_2 = m \nconc m_1 ~\vee~ m_2 = m_1 \nconc m) \wedge\\ \knows{A}{m_2})\\
~~\vee\\
(\exists m_1, m_2: m = m_1 \nconc m_2 ~\wedge~ \knows{A}{m_1}  \wedge\\ \knows{A}{m_2})\\
~~\vee\\
(\exists k, k^{-1}: \know{A}{k^{-1}} ~\wedge~ \knows{A}{Enc(k, m)})\\
~~\vee\\
(\exists k, k^{-1}: \know{A}{k} ~\wedge~ \knows{A}{Sign(k^{-1}, m)})\\
~~\vee\\
(\exists k, m_1: m = Enc(k, m_1) ~\wedge~ \knows{A}{m_1} ~\wedge\\ \know{A}{k})\\
~~\vee\\
(\exists k, m_1: m = Sign(k, m_1)  \wedge \knows{A}{m_1} \wedge \\ \know{A}{k})
\end{array}
\]
Subsequently, we add  axioms that describe relations between the predicates $know /knows$ and the predicate describing 
that a component may
eventually output an expression.

\begin{ax}
\label{ax:knows}
For any component $C$ and for any secret $m \in KS$ (or expression $e \in \nfst{Expression}$), 
the following equations hold:
\[
	\begin{array}{l}
		\forall m \in KS:~ \eout{C}{m} ~\equiv~ (m \in KS_C) ~\vee~ \know{C}{m}\\
		\forall e \in \nfst{Expression}:\\ \eout{C}{e} ~\equiv~ (e \in \nfst{KS_C}) ~\vee~ \knows{C}{e}
	\end{array}
\]
\end{ax}

\begin{ax}
\label{ax:knows_empty}
For any component $C$ and for an empty expression $\nempty \in \nfst{Expression}$, 
 holds:\\
$\forall C:~ \knows{C}{\nempty} = \ntrue
$.
\end{ax}

For the collection (more than 50 Isabelle lemmas) of propositions and theorems on the properties of the predicates $know /knows$, we would like to refer to~\cite{spichkova2014compositional}. 


\section{\uppercase{TLS Protocol}}
\label{sec:tls}

\noindent 
To demonstrate the feasibility and usability of our approach, we specified a variant
of the TLS protocol\footnote{TLS 
(Transport Layer Security) is the successor of the Internet security protocol
SSL (Secure Sockets Layer).} in \focust and discuss the
features of the specification templates that were introduces to increase the readability of the language.  
After that we present the formal analysis of the protocol. 

\begin{figure}[ht!]
  \centering
   { \includegraphics[scale=0.75]{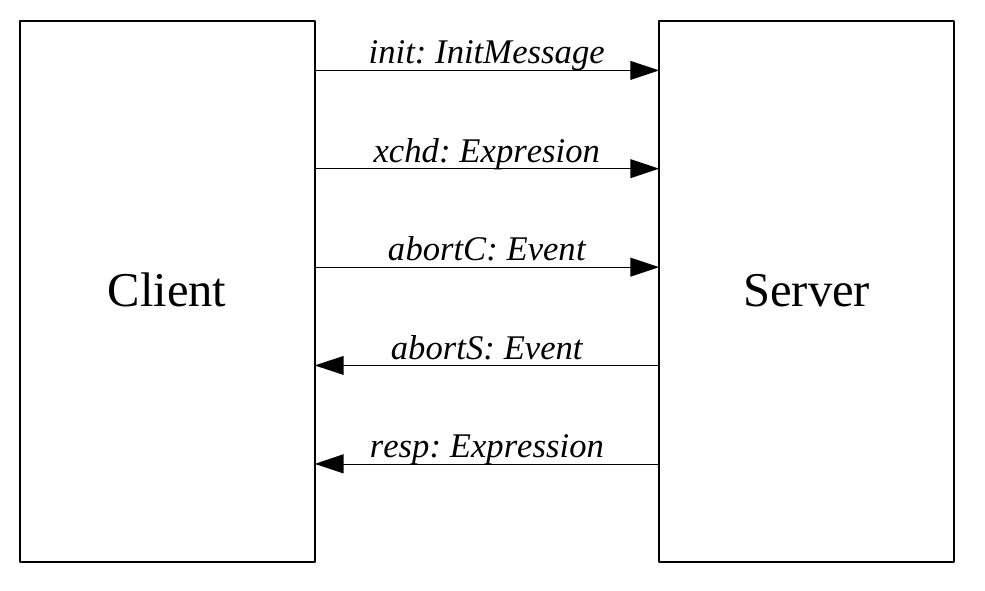}}
  \caption{Protocol of TLS}
  \label{fig:tls1}
 \end{figure}

The goal of TLS  is to let a client send a secret over an untrusted communication
link to a server in a way that provides secrecy and server authentication, by
using symmetric session keys.  
Let us recall the general idea of the handshake protocol of TLS, cf. Figure~\ref{fig:tls1}. 
The protocol has two participants, \emph{Client} and \emph{Server}, that  
 are connected by an Internet connection. 
We used the following auxiliary data types: 
\begin{itemize}
\item 
$Obj = \{ C, S \}$ to represent participants names ($C$ for the \emph{Client} and $S$ for the \emph{Server}),
\item
$StateC = \{ st0, st1, st2 \}$ to represents the states of the \emph{Client},
\item
$StateS = \{initS, waitS, sendS1, sendS2\}$ 
to represent the states of the \emph{Server},
\item
$Event  = \{event\}$  to represent events of message sending  
(e.g.,\ an abort message or an acknowledgement), and 
\item
$InitMessage = im(ungValue \in \secret, key \in \keys,~ msg \in Expression)$
   to represent the event of protocol initiation by the \emph{Client}.
\end{itemize}

\begin{figure}[ht!]
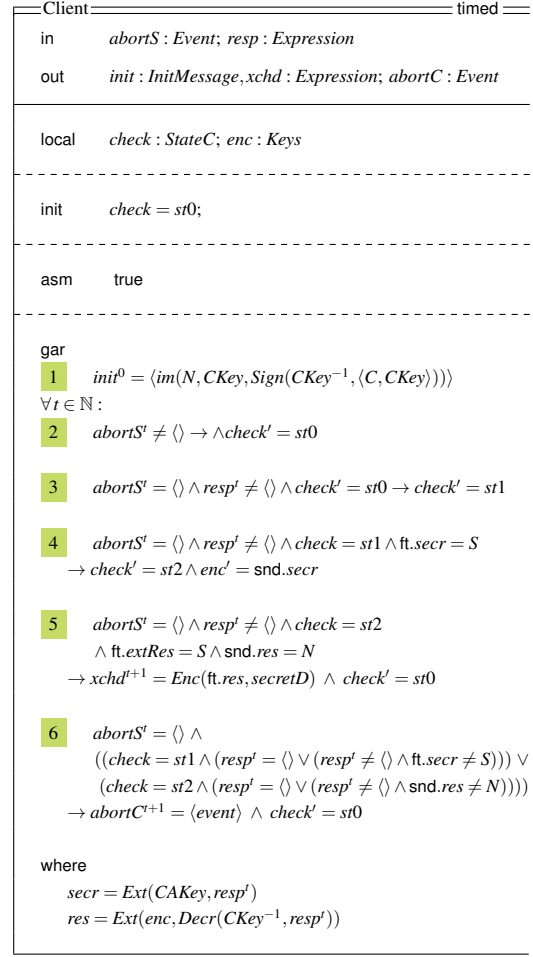

  \centering
{\scriptsize
\begin{spec}{Client}{timed}
\InOut{abortS: Event; resp: Expression}
{init: InitMessage, xchd: Expression; abortC: Event}
\nlocal
       check: StateC; enc: Keys 
\ninit  check = st0; 
\zeddashline
\tab{asm}~  \ntrue
\zeddashline
\tab{gar}\\
\cbox{1}  \t1 
\ti{init}{0} = \angles{im(N, CKey, Sign (CKey^{-1}, \angles{C, CKey}))}\\
\forall t \in \Nat:\\
\cbox{2}
\t1  \ti{abortS}{t} \neq \nempty  \to 
  \wedge check' = st0\\
~\\
\cbox{3} 
\t1  \ti{abortS}{t} = \nempty \wedge \ti{resp}{t} \neq \nempty \wedge check' = st0  \to %
check' = st1\\
~\\
\cbox{4} 
\t1
 \ti{abortS}{t} = \nempty \wedge \ti{resp}{t} \neq \nempty \wedge check = st1   \wedge \nft{secr} = S \\
\t1 \to   
check' = st2 \wedge enc' = \snd{secr} \\  
~\\
\cbox{5} 
\t1
 \ti{abortS}{t} = \nempty \wedge \ti{resp}{t} \neq \nempty \wedge check = st2 \\
 \t2 \wedge~
			   \nft{extRes}= S  \wedge 
			  \textsf{snd}.res  = N \\
\t1 \to   
\ti{xchd}{t+1} =  Enc(\nft{res}, secretD) ~  \wedge~ check' = st0\\    
~\\
\cbox{6}  
\t1
 \ti{abortS}{t} = \nempty  ~\wedge \\
 \t2  ((check = st1 \wedge (\ti{resp}{t} = \nempty \vee (\ti{resp}{t} \neq \nempty  \wedge \nft{secr} \neq S))) ~\vee\\
 \t2  ~(check = st2 \wedge (\ti{resp}{t} = \nempty \vee  (\ti{resp}{t} \neq \nempty  \wedge \textsf{snd}.res  \neq N)))) \\
\t1 \to  \ti{abortC}{t+1} = \angles{event} ~\wedge~
       check' = st0\\	
~\\
\nwhere\\
\t1 secr = Ext(CAKey, \ti{resp}{t})\\
\t1 res = Ext(enc,  Decr(CKey^{-1}, \ti{resp}{t}))		         
\end{spec}
}
  \caption{\focust\ specification of the \emph{Client} component}
  \label{fig:client1}
 \end{figure}

\emph{Client} initiates the protocol by sending the message that contains 
an unguessable value $N \in \secret$, its the public key $K_C$, and 
a sequence $\angles{C, CKey}$ of its name and its public key signed by its secret key $CKey^{-1}$. 
\emph{Server} checks whether the received public key matches to the second element of the signed sequence. 
If that is the case, it returns to the \emph{Client} 
the received unguessable value $N$, 
an encryption of a sequence $\angles{genKey, N}$ (signed by its secret key $SKey^{-1}$) using the received public key, and 
a sequence $\angles{S, SKey}$ of its name and its public key,  signed using the secret key $CAKey^{-1}$ of the certification authority. 
After that, \emph{Client} checks whether the certificate is actually for $S$ and the correct $N$ is returned. 
If that is the case, it sends the secret  value \emph{secretD} 
encrypted with the received session key $genKey$ to the \emph{Server}. 
If any of the checks fail,
the respective protocol participant stops the execution of the protocol by sending
an abort signal. 
Figures~\ref{fig:client1} and \ref{fig:server1} present the \focust\ specification of \emph{Client} and \emph{Server} components  respectively. 

For the  corresponding representation in \isah we would like to refer to the technical report. 
We continue with the discussion how our approach can be used to demonstrate a security
flaw in the TLS variant introduced above, as well as how to correct it.

\begin{figure}[ht!]
  \centering
{\scriptsize
\begin{spec}{Server}{timed}
\InOut{init: InitMessage; abortC: Event; xchd: Expression}
{resp: Expression; abortS: Event}
\tab{\ulocal} ~ stateS \in StateS; kValue \in Keys; uValue \in Secret
\zeddashline
\tab{\uinit} stateS = initS
\zeddashline
\tab{asm}~\msg{1}{init} ~\wedge~ \msg{1}{xchd}\\
\zeddashline
\tab{gar}\\
\forall t \in \Nat:\\
\cbox{1} 
 \t1 
\ti{abortC}{t} \neq \nempty  \to
	   stateS' = initS  \\ 
~\\
\cbox{2} 
\t1 
\ti{abortC}{t} = \nempty ~\wedge~ stateS = initS ~\wedge~ \ti{init}{t} \neq \nempty ~\wedge\\
\t2 \snd{Ext(\angles{key(\fti{init}{t}), msg(\fti{init}{t})})} \neq  key(\fti{init}{t})\\
\t1 \to  %
\ti{abortS}{t+1} = \angles{event}\\
~\\
\cbox{3}  
\t1 
\ti{abortC}{t} = \nempty ~\wedge~ stateS = initS ~\wedge~ \ti{init}{t} \neq \nempty ~\wedge\\
\t2 \snd{Ext(\angles{key(\fti{init}{t}), msg(\fti{init}{t})})} =  key(\fti{init}{t})\\
\t1 \to    \ti{resp}{t+1}= \angles{ungValue(\fti{init}{t})} \\
	  \t2   \wedge~  stateS' = sendS1  ~\wedge~  uValue' = ungValue(\fti{init}{t}) \\
	  \t2 \wedge~  kValue' = key(\fti{init}{t}) \\
~\\
\cbox{4} 
\t1 
\ti{abortC}{t} = \nempty ~\wedge~ stateS = sendS1  \\
\t1 \to    \ti{resp}{t+1}= Sign(CAKey^{-1}, \angles{S, SKey})  
          ~ \wedge~ stateS' = sendS2 \\ 
~\\
\cbox{5} 
\t1 
\ti{abortC}{t} = \nempty ~\wedge~ stateS = sendS2  \\
\t1 \to    \ti{resp}{t+1}= Enc(kValue,  Sign(SKey^{-1}, \angles{genKey, uValue})) \\
	     \t2 \wedge~  stateS' = waitS \\ 
\end{spec}
}
  \caption{\focust\ specification of the \emph{Server} component}
  \label{fig:server1}
 \end{figure}
 
 We specify every component using assumption-guarantee-struc\-tu\-red templates.
This helps avoiding the omission of unnecessary assumptions about the system's environment since a specified component is required to fulfil the guarantee only if its environment behaves in accordance with the assumption.
We also use in 
\focust\ so-called \emph{implicit else-case} constructs: 
 if a variable is not listed in the guarantee part of a transition, it implicitly keeps its current value.
An output stream not mentioned in a transition will be empty.

 Without these extensions, the specification of the components would become less readable, 
as the core properties would be lost in a huge set of properties that might be specified implicitly.
For example, the specification of the \emph{Client} would require 4 additional properties, such as
\begin{itemize} 
\item
$\ti{xchd}{0} = \nempty$,
\item
$\ti{abortC}{0} = \nempty$,
\item
$\ti{init}{t+1} = \nempty$,
\item
$ \ti{abortS}{t} = \nempty \wedge \ti{resp}{t} = \nempty \wedge check = st0 
 \to 
\ti{abortC}{t+1} = \nempty   \wedge \ti{xchd}{t} = \nempty   \wedge check' = st0$
\end{itemize}
Thus, the corresponding \Focus\ specifications of  \emph{Client} and \emph{Server} would have 10 and 9 properties respectively, in contrast to 6 and 5 properties we have in the \focust\ version. 
Even when for a small example like TLS the difference might not look huge, it scales dramatically for large systems.

Moreover, the specifications of the properties would be more complicated without these optimisations.
For example, the 4th property of \emph{Server} in the \Focus\ version would be
\[\begin{array}{l}
\ti{abortC}{t} = \nempty ~\wedge~ stateS = sendS1\\ 
\to    \ti{resp}{t+1}= Sign(CAKey^{-1}, \angles{S, SKey})  
          \wedge~ \\
\t1 stateS' = sendS2 ~\wedge~ uValue' = uValue ~\wedge~  \\
\t1 kValue' = kValue
	    \wedge~   \ti{abortS}{t+1} = \nempty
\end{array}    
\]
In \focust\,  also do not require to introduce auxiliary variables explicitly:
The data type of an unintroduced variable is universally quantified in the specification such that it can be used with any data value.


\subsection{Security Analysis}
\label{sec:SecurityAnalysis}

\begin{figure*}[!ht]
 \centering
  \includegraphics[scale=0.75]{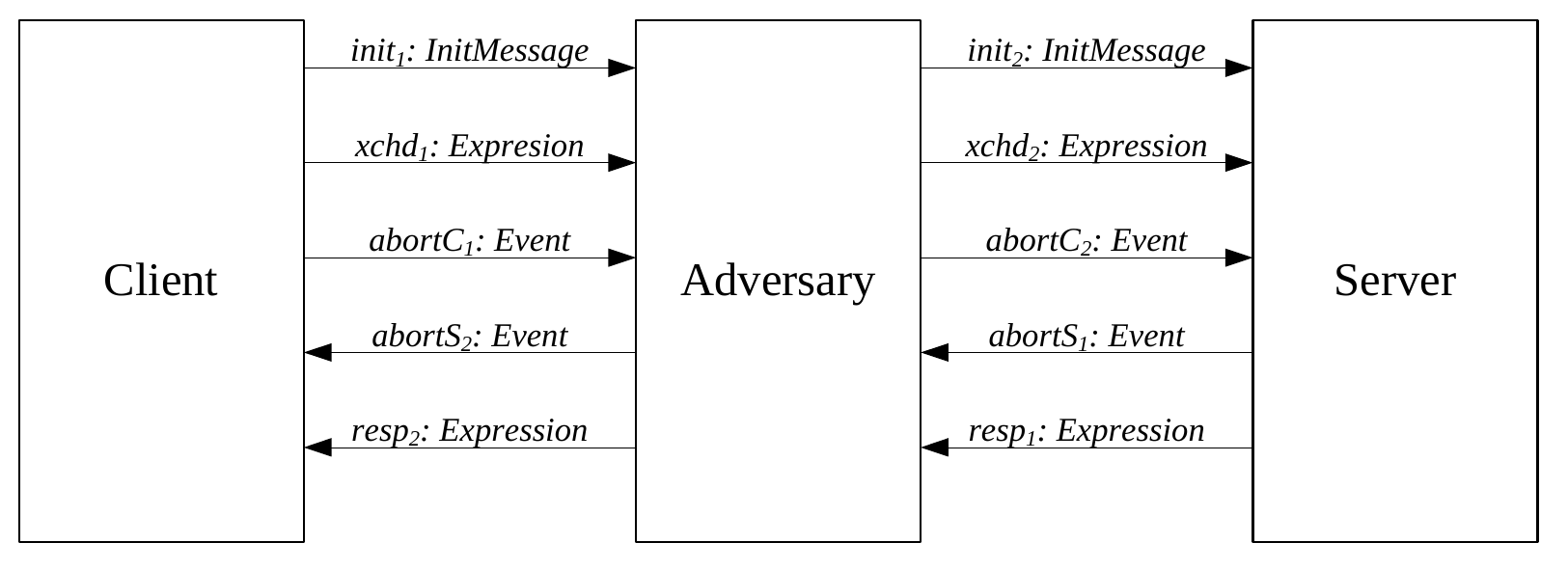}
  \caption{Protocol of TLS: Situation with an \emph{Adversary} component involved}
  \label{fig:tls2}
 \end{figure*}

In this section, we use our approach to demonstrate a security
flaw in the TLS variant introduced above, and how to correct it.
Let us  assume a composite component $P = Client \otimes Server$. 
To show that $P$ does not preserve the secrecy of $secretD$,  $secretD \in KS$, we need to find an adversary component $Adversary$ with $\sse{I_{Adversary}}{O_P}$ such 
that  
\begin{itemize}
\item 
$\knows{Adversary}{m}$ holds with regards to the composition, and 
\item
$m$ does not belong to the set of private keys of $Adversary$ or to the set of unguessable values of $Adversary$,
\end{itemize} 
This can be formalised as the below statement:
\begin{equation}
\begin{array}{l}
	\exists Adversary:  \\
	\sse{I_{Adversary}}{O_P} ~\wedge~ m \notin KS_{Adversary} \\
	~\wedge~ \knows{Adversary}{m}
\end{array}
\label{eq:Adv1}
\end{equation}
\noindent
This means, we have to analyse a possible composition $Client \otimes Adversary \otimes Server$, cf. Figure~\ref{fig:tls2}.

The protocol assumes that there is a secure (wrt.\ integrity) way for the client to obtain
the public key \emph{CAKey} of the certification authority, and for the server to obtain
a certificate 
\[Sign(CAKey^{-1}, \angles{S, SKey})
\]
signed by the certification authority that contains
its name and public key. 
For an arbitrary process $Z$, an adversary may also have access to 
\begin{itemize}
\item \emph{CAKey}, 
\item
$Sign(CAKey^{-1}, \angles{S, SKey})$, and 
\item 
$Sign(CAKey^{-1}, \angles{Z, ZKey})$.
\end{itemize}

Consider the  \focust\ specification of the component \emph{Adversary} presented in Figure~\ref{fig:adv}. This component is weakly causal: we assume that the adversary does not delay any message. 
We used in this specification an auxiliary data type\\  

$AdvStates = \{initA, sendA1, sendA2\}$.

\begin{figure}[!ht]
 \centering
{\scriptsize
\begin{spec}{Adversary}{timed}
\InOut{\begin{array}{l}
				abortC_1, abortS_1: Event; xchd_1, resp_1: Expression; \\init_1: InitMessage 
				\end{array}}
     {\begin{array}{l}
     abortC_2, abortS_2: Event;  xchd_2, resp_2: Expression; \\init_2: InitMessage
     \end{array}}
\tab{\ulocal} ~ aCKey, aSKey, aKey \in \keys; stateA \in AdvStates
\zeddashline
\tab{asm}~ 
			\msg{2}{resp_1} ~\wedge~ 
			 \msg{1}{xchd_1}\\
\zeddashline
\tab{gar}\\
\forall t \in \Nat:\\
\cbox{1}  \t1
\ti{abortC_2}{t} = \ti{abortC_1}{t}\\%
\cbox{2}  \t1
\ti{abortS_2}{t} = \ti{abortS_1}{t}\\
\cbox{3}  \t1 
\ti{xchd_2}{t} =  \ti{xchd_1}{t}\\
~\\
\cbox{4}  \t1 
\ti{init_1}{t} \neq \nempty  \\
\to  aCKey' = key(\fti{(init_1)}{t}) ~\wedge\\
		 \t1\ti{init_2}{t} =  \langle im(ungValue(\fti{(init_1)}{t}), AKey,   Sign (AKey^{-1}, \angles{C, AKey}))\rangle\\
~\\
\cbox{5} %
\t1
\ti{resp_1}{t} \neq \nempty \wedge stateA = initA \\ \t1 \to 
   	  stateA' = sendA1 ~\wedge~ %
	 \ti{resp_2}{t} = \ti{resp_1}{t} \\
~\\
\cbox{6}  \t1 
\ti{resp_1}{t} \neq \nempty \wedge stateA = sendA1 \\
\t1 \to  stateA' = sendA2 ~\wedge~  aSKey' = \snd{Ext(CAKey, \ti{resp_1}{t})} ~\wedge\\
	\t2 %
	 \ti{resp_2}{t} = \ti{resp_1}{t}\\
~\\
\cbox{7}  \t1 
\ti{resp_1}{t} \neq \nempty \wedge stateA = sendA2 \\
\t1 \to  stateA' = initA ~\wedge~ \\%
	\t2 aKey = \uft.{Ext(aSKey,  Decr(AKey^{-1}, \ti{resp_1}{t}))}\\
	\t2  \ti{resp_2}{t} =  Enc(aCKey, Decr(AKey^{-1}, \ti{resp_1}{t}))\\
\end{spec}
}
  \caption{\focust\ specification of the \emph{Adversary} component that fulfils statement (\ref{eq:Adv1})}
  \label{fig:adv}
 \end{figure}

The value $genKey \in \keys$ is a symmetric session key generated by the server: $genKey^{-1} = genKey$. 
This implies that 
\[
\knows{Adversary}{genKey}
\]
holds if and only if 
\[
\knows{Adversary}{genKey^{-1}}
\] 
holds. 
Thus, if the adversary knows the value of $genKey$ it also knows the value of $genKey^{-1}$. 
If we trace its knowledge base as its evolves in interaction with the
protocol components, we get that $Adversary$ will know the secret $secretD$ at the time unit 4.

Translating the \focust\ specifications to Isabelle/HOL, 
we can prove formally that the security
flaw exists. These proof (together with protocol component specifications and auxiliary lemmas) takes 1,5 klop (thousands lines of proofs). 
This also allows us to demonstrate the correctness of
syntactic interfaces for specified system components automatically.

In this paper we present only the main lemma which says that the during the 4th time unit the secret data $secretD$ will be send to the adversary by the $Client$ component and no abort-signal will be produced: 
For further details we would like to refer to the Isabelle/HOL-theories we uploaded to the Archive of Formal Proofs~\cite{spichkova2014compositional}. %

\subsection{Fixing the Security Weakness}
\label{sec:Correction}

To fix the security weakness, we need to change the protocol: the client must find out the situation, 
where an adversary try to get the secret data.
Thus, we need to correct the specification of the server in such a way that the client will
know with which public key the data was encrypted at the server, and this information 
must be received by the client without any possible changes by the adversary. 
The only part of the messages from the server which cannot be changed by the adversary 
is the result of the signature creation -- the adversary does not know the secret key $SKey^{-1}$
and cannot modify the signature or create a new one with modified content. 
Therefore, we add  the 
public key received by the server to the content $\angles{genKey, N}$ of the  signature.
If there is not attack, this will be $CKey$, in the attack scenario explained above, it would be $AKey$.
Accordingly, in the specification of the \emph{Server}, we change the value of $\ti{resp}{t+1}$  in the 5th formula to the following one:
\[ \begin{array}{l}
Enc(key(\fti{init}{t}),~ \\
~~~~~~~Sign(SKey^{-1},~ \\
~~~~~~~~~~~~~~\angles{genKey, ungValue(\fti{init}{t}), {key(\fti{init}{t})}}))
\end{array}
\]
\noindent
Also, correspondingly we add a new conjunct to the condition for the correct data receipt
in the specification of the client:
\[\begin{array}{l}
\textsf{trd}.Ext(\snd{Ext(CAKey, \tti{resp}{t})},\\
~~~~~~~~~~~~~~~~~ Decr(CKey^{-1}, \sti{resp}{t}))\\
 = CKey 
 \end{array}
\]
If we trace the knowledge base of the adversary $Adversary$ considered above,
the secret is not leaked, the transmission will be aborted by the client on the 4th time unit.

We omit here the complete presentation of the \Focus and \isah specification of the corrected components $Client$ and $Server$, because they have only the minor changes vs.\ the specification presented in the previous section. 
The \isah proof takes also about 1,5 klop, but the most number of lemmas can be reused from the uncorrected version.


\section{\uppercase{Conclusions}}
\label{sec:conclusions}

\noindent 
In this paper, we present a theory that allows (1)~to specify distributed
systems formally, (2) to verify their cryptographic wrt. composition properties, and
(3)  to demonstrate the correctness of
syntactic interfaces for specified system components automatically. 
The theory is based on the \focust\ formal language, and the verification is conducted using the theorem prover  Isabelle/HOL. 

The feasibility of this approach was demonstrated by specifying and formally analysing a variant of the Internet security protocol TLS, which is a typical example from the domain of crypto-based systems.
We analysed the protocol using both  paper-and-pencil proofs and the automated verification with Isabelle/HOL.
The analysis revealed a security flaw in the initial version of TLS specification.   
The protocol specification was harden according the proposed approach.

\bibliographystyle{apalike}
{\small

}

\vfill
 
\end{document}